\documentclass{article}

\PassOptionsToPackage{numbers, compress}{natbib}

 \usepackage[dblblindworkshop, final]{neurips_2025}

\workshoptitle{Machine Learning and the Physical Sciences}

\usepackage[utf8]{inputenc} 
\usepackage[T1]{fontenc}    
\usepackage{hyperref}       
\usepackage{url}  
\usepackage{booktabs}       
\usepackage{amsfonts}       
\usepackage{nicefrac}       
\usepackage{microtype}      
\usepackage{xcolor}         

\usepackage{amsmath}
\usepackage{todonotes}
\usepackage{enumitem}

\title{Implicit Augmentation from Distributional Symmetry in Turbulence Super-Resolution}

\author{%
  Julia Balla\textsuperscript{*}\\
  \texttt{jballa@mit.edu}
  \And
  Jeremiah Bailey\textsuperscript{*}\\
  \texttt{jeremiah.bailey@bison.howard.edu}
  \And
  Ali Backour\\
  \texttt{abackour@mit.edu}
  \And
  Elyssa Hofgard\\
  \texttt{ehofgard@mit.edu}
  \And
  Tommi Jaakkola\\
  \texttt{jaakkola@mit.edu}
  \And
  Tess Smidt\textsuperscript{$\dagger$}\\
  \texttt{tsmidt@mit.edu}
  \And
  Ryley McConkey\textsuperscript{$\dagger$}\\
  \texttt{rmcconke@mit.edu}
  \AND
  {\normalfont\parbox{\linewidth}{\centering Massachusetts Institute of Technology, Cambridge, MA 02139\\
  \textsuperscript{*,$\dagger$} Equal contribution\\
  }}
}

\begin{document}

\maketitle

\begin{abstract}
The immense computational cost of simulating turbulence has motivated the use of machine learning approaches for super-resolving turbulent flows. A central challenge is ensuring that learned models respect physical symmetries, such as rotational equivariance. We show that standard convolutional neural networks (CNNs) can partially acquire this symmetry without explicit augmentation or specialized architectures, as turbulence itself provides \emph{implicit} rotational augmentation in both time and space. Using 3D channel-flow subdomains with differing anisotropy, we find that models trained on more isotropic mid-plane data achieve lower equivariance error than those trained on boundary layer data, and that greater temporal or spatial sampling further reduces this error. We show a distinct scale-dependence of equivariance error that occurs regardless of dataset anisotropy that is consistent with Kolmogorov’s local isotropy hypothesis. These results clarify when rotational symmetry must be explicitly incorporated into learning algorithms and when it can be obtained directly from turbulence, enabling more efficient and symmetry-aware super-resolution.
\end{abstract}

\section{Introduction}

Super-resolution (SR) with machine learning has become a promising method to augment numerical simulations of turbulence, by boosting the effective resolution of expensive calculations \cite{Duraisamy2019}. In this setting, convolutional neural networks (CNNs) have been widely applied to reconstruct velocity and vorticity fields from coarse inputs \citep{Fukami_Fukagata_Taira_2019, Liu2020, FukamiFukagataTaira_2024, Pang2024}. Complementary approaches span models that incorporate temporal coherence and dynamics-aware training objectives \citep{Fukami_Fukagata_Taira_2021, Page_2025}, as well as generative methods such as generative adversarial networks (GANs) \citep{Nista2024} or diffusion models that reproduce realistic spectra and scaling laws \citep{Shu2023,WhittakerNairLivescuChertkov_2024, Fan2025}. 

An important dimension of physical consistency is the treatment of symmetries. In turbulent flows, the velocity field $\mathbf{U}(\mathbf{x}, t)$ is a random vector field whose statistics may exhibit distributional symmetries, i.e., invariance of the probability law under a group action, $p(\mathbf{U}) = p(g \cdot \mathbf{U})$ for transformations $g$ such as translations (\emph{homogeneity}) or rotations/reflections (\emph{isotropy}). \emph{Kolmogorov’s local isotropy hypothesis} further asserts that when turbulence is strong enough for small eddies to form without being damped by viscosity, small-scale motions approach isotropy even when large scales are anisotropic \citep{Kolmogorov1991,Pope_2000}. A formal treatment of statistical isotropy is provided in Appendix \ref{app:definitions}. 

\textit{Equivariance} is often an important inductive bias in machine learning for physics, as it ensures that models respect the symmetries of the underlying system \cite{ai4sci}. In short, an equivariant model $f$ satisfies $f(g \cdot \mathbf{U}) = g \cdot f(\mathbf{U})$, so that outputs transform consistently with the inputs. Several strategies have been proposed to incorporate equivariance into SR. Architectural approaches impose symmetry exactly through group-equivariant convolutions or neural operators \citep{HelwigZhangFuEtAl_2023,XuHanLouEtAl_2024}, while loss-based approaches regularize models to transform consistently without altering their backbone \citep{BaiEtAl_2025, Raissi2019}. Relaxed group convolutions have been used to probe isotropy-breaking in turbulence by permitting departures from exact equivariance, highlighting its scale dependence \citep{Wang2024}. \citet{YasudaOnishi_2023} showed that CNNs learn rotation equivariance only when the coarsening operator commutes with rotations, sharpening the question of when equivariance can emerge implicitly from data.

These considerations motivate the central question of this study: to what extent does turbulence itself provide the rotational augmentation required for learning equivariance? We investigate this question by analyzing the role of statistical isotropy in determining whether rotational equivariance can be learned implicitly from data or must be imposed through explicit transformations. To this end, we compare models trained on channel-flow subdomains with differing anisotropy, contrasting boundary-layer regions against the more isotropic mid-plane. 
Our results demonstrate that:
\vspace{-2pt}

\begin{itemize}[leftmargin=*]
    \item Imposing explicit rotational augmentation during training improves generalization to unseen test data, indicating that rotational symmetry is a useful inductive bias for turbulence SR.  
    \item Increasing the temporal and spatial domain spanned by the training data increases its statistical isotropy, enabling models to acquire more rotational equivariance without explicit augmentation.
    \item Equivariance error exhibits a distinct scale dependency, consistent with the stronger small-scale isotropy predicted by Kolmogorov’s hypothesis.  
\end{itemize}
\vspace{-2pt}
Collectively, these findings establish turbulence as a setting where understanding the interplay between data symmetries and model design is essential for physically grounded learning.
\section{Methodology}

\begin{figure}
  \centering
  \includegraphics[scale=.51]{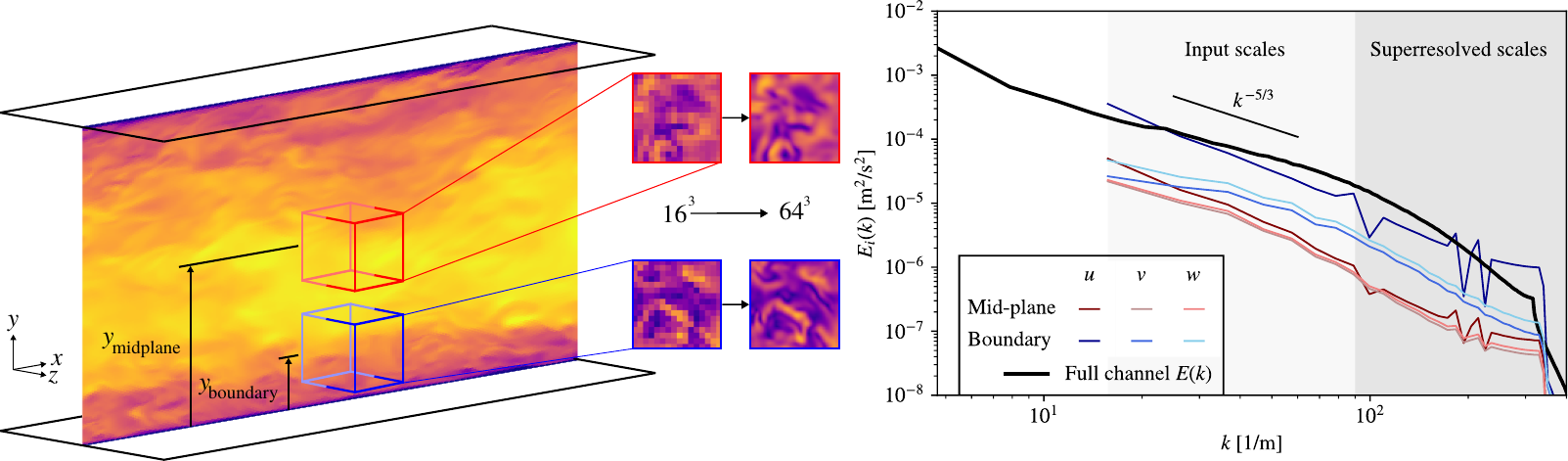}
  \caption{\textbf{Box placement and energy spectra for turbulent channel flow.} The superresolution task involves filling in unresolved scales of turbulence. The energy spectra (right) show significant anisotropy in the turbulent boundary layer near the channel walls. Note that in the present study, we take $k$ as the angular wavenumber.}\label{fig:domain_spectra}
\end{figure}
Our study aims to test whether CNNs can acquire rotational equivariance implicitly from being trained to perform SR on turbulence data. The input to the CNNs are turbulent velocity fields discretized on a 3D Cartesian grid with each grid cell storing the velocity components ($u$, $v$, $w$) as the input channel dimension. To assess whether models preserve rotational symmetries, we evaluate their \textit{equivariance error}. Given an input velocity field $\overline{\mathbf{U}}$ and a model $f$, for any rotation $g \in G$, the absolute equivariance error is defined pointwise as $\mathcal{E}(\overline{\mathbf{U}}; g) \;=\; \big\lVert f(g \cdot \overline{\mathbf{U}}) - g \cdot f(\overline{\mathbf{U}}) \big\rVert$, where $g \cdot \overline{\mathbf{U}}$ denotes the rotated input and $g \cdot f(\overline{\mathbf{U}})$ the rotated model output. Averaging over all group elements and $N$ samples yields the overall equivariance error 
\begin{equation}
\overline{\mathcal{E}} \;=\; \frac{1}{|G|N} \sum_{g \in G} \sum_{n=1}^N \mathcal{E}(\overline{\mathbf{U}}_n; g).
\end{equation}
We evaluate equivariance error over the discrete octahedral group $O$ (rotations without inversions). Further discussion is provided in Appendix~\ref{app:rotation_symmetry}.

\subsection{Model architecture}
We employ a compact multi-scale convolutional super-resolution network that upsamples a low-resolution velocity field volume $\overline{\mathbf{U}} \in \mathbb{R}^{3\times D \times H \times W}$ to the target high-resolution $\mathbf{U} \in \mathbb{R}^{3\times sD \times sH \times sW}$. Upsampling by factor $s$ is implemented as a sequence of resize-then-refine stages, one for each factor of 2. At each stage, the input is upsampled through trilinear interpolation and passed through two convolutional layers. A final convolution projects the features to 3 output channels, yielding the super-resolved prediction. The SR model is trained to minimize the mean absolute error (MAE) loss between the ground truth and predicted high resolution fields, which has been shown to better preserve perceptual quality and reduce oversmoothing compared to mean squared error (MSE) loss in image restoration tasks \citep{Zhao2017}. See Appendix \ref{app:training_params} for details on training hyperparameters.

\subsection{Dataset}

We evaluate models on 3D channel-flow subdomains drawn from direct numerical simulation found in the Johns Hopkins Turbulence Database \cite{Li_2008}. We subsample the available data into 150 evenly spaced timesteps for the experiments in our study. To probe the effect of anisotropy, we compare (i) boundary-layer regions near the channel wall, where turbulence is strongly anisotropic, against (ii) mid-plane subdomains, where it is closer to isotropic (see Figure~\ref{fig:domain_spectra} for the component-wise energy spectra). To test temporal data augmentation, we sample from a time series of a single subdomain at a fixed $y$-coordinate (see Figure~\ref{fig:domain_spectra}). The training set is sampled from the first 100 timesteps, while validation and test sets consist of the following 30 and 20 timesteps. To test spatial data augmentation, we add randomly sampled (without overlap) subdomains at a fixed $y$-coordinate and timestep. 

Subdomain sizes were carefully selected in order to verify the implicit data augmentation hypothesis. The subdomains should contain inertial range length scales, where turbulence begins to cascade towards isotropy. As shown in Figure~\ref{fig:domain_spectra}, a significant portion of the inertial range (where $E(k) \sim k^{-5/3}$) of the energy cascade is captured within each subdomain. The low-resolution inputs are obtained by applying a box filter and downsampling the high-resolution fields by a scaling factor of $s=4$, thereby truncating the available lengthscales in the input vector fields. All code and data used in our experiments is available at \href{https://github.com/atomicarchitects/turbulence-implicit-augmentation}{https://github.com/atomicarchitects/turbulence-implicit-augmentation}.

\section{Results}


\begin{figure}[t]
    \centering
    \includegraphics[scale=0.542]{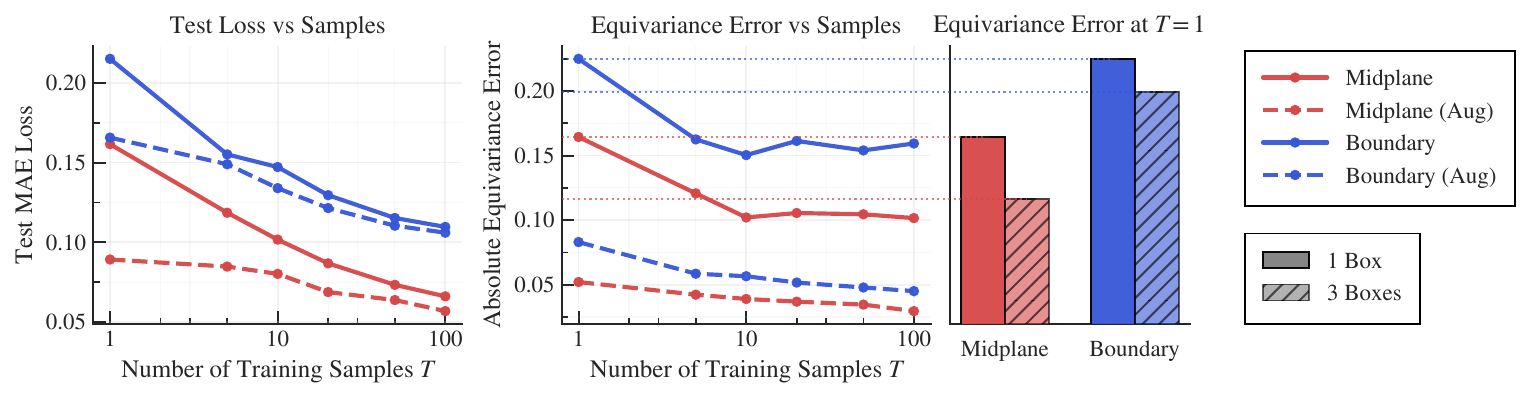}
    \caption{\textbf{Generalization and equivariance error under varying sampling and data augmentation.} Left: Test MAE loss decreases with more training samples $T$, with dashed curves showing the added benefit of rotational augmentation. Middle: Absolute equivariance error also declines with $T$, with mid-plane models (red) consistently below boundary models (blue). Right: Bar plots at $T=1$ highlight reduced error when training on larger domains (lighter shaded bars).}
    \label{fig:scaling-samples}
\end{figure}

\begin{figure}[t]
    \centering
    \includegraphics[scale=0.6]{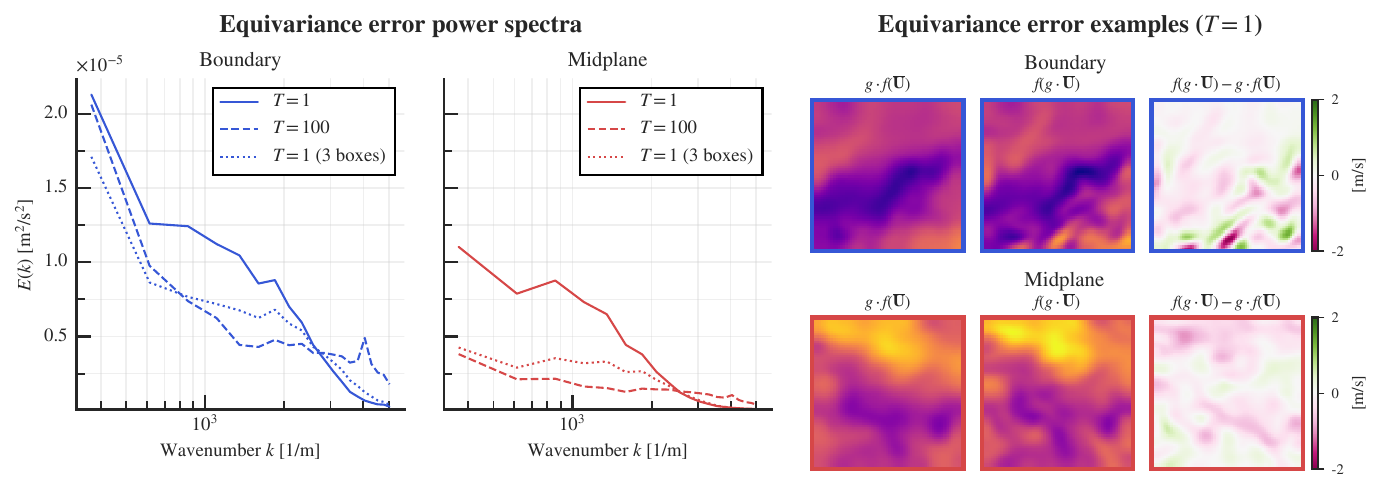}
    \caption{\textbf{Equivariance error across scales.} (a) (Left) Power spectra show larger, scale-dependent errors in boundary-layer turbulence, while mid-plane and larger spatio-temporal domains reduce errors through greater isotropy. (Right) Example $x$-component error fields (in standardized velocity units) at mid-$z$ slice with $T=1$ illustrate smaller residuals in isotropic regimes.}
    \label{fig:ee_spectra}
\end{figure}

\textbf{Implicit augmentation over time.} We first assess how the number of timesteps used for training affects super-resolution accuracy and rotational equivariance. Training sets consist of the first $T \in \{1, 5, 10, 20, 50, 100\}$ timesteps of each simulation.  To test for implicit rotational coverage, we compare models trained on raw data against those trained with explicit octahedral augmentation, in which each training snapshot is randomly rotated at every epoch. Figure~\ref{fig:scaling-samples} shows results for the anisotropic boundary layer and the more isotropic mid-plane. Test MAE decreases steadily with $T$, while equivariance error drops quickly and saturates after only a few snapshots. Explicit augmentation reduces both MAE and equivariance error, with the strongest benefit in low-data regimes and in anisotropic boundary data. By contrast, mid-plane models benefit less, consistent with stronger implicit augmentation from isotropy. We further characterize these effects by computing the Fourier power spectrum of equivariance error fields (Figure~\ref{fig:ee_spectra}). High-wavenumber modes exhibit consistently lower error across all training conditions, confirming that small scales act as a natural source of rotational consistency. Enlarging the training set in time or applying explicit augmentation primarily reduces error at intermediate scales, while all models converge in the dissipative range.

\textbf{Implicit augmentation over space.} Kolmogorov’s hypothesis suggests that increasing the spatial domain size should capture more isotropic small-scale motions. To test this, we fix the training set to a temporal snapshot $(T=1)$ and increase the number of boxes extracted from the channel flow at a fixed $y$-coordinate. In particular, for the original boundary and mid-plane boxes, we randomly sample two additional boxes at the same respective $xz$ plane such that the expected degree of anisotropy in each box is maintained (by $xz$-plane homogeneity of this flow). As shown in Figures~\ref{fig:scaling-samples} and~\ref{fig:ee_spectra}, the larger spatial domain consistently yields lower equivariance error, particularly at intermediate and large scales. Notably, training on three boxes from a single snapshot achieves an equivariance error comparable to training on 100 sequential snapshots in the mid-plane case. This reflects the fact that temporally adjacent snapshots are strongly correlated, whereas spatially distinct boxes supply more diverse and less redundant samples. Future work will investigate how temporally correlated snapshots affect a model's ability to learn equivariance~\cite{Sardar2024}. Additionally, our results extend the single-snapshot SR results of Fukami et al.~\cite{FukamiFukagataTaira_2024} by showing that rotational equivariance can also be partially learned from a single snapshot of turbulence. 

\section{Conclusion}

We have shown that statistical isotropy in turbulence acts as an implicit form of data augmentation, enabling convolutional models to acquire rotational equivariance without explicit enforcement. By analyzing equivariance error across both temporal and spatial sampling, we demonstrated that turbulence provides scale-dependent distributional symmetry: small scales consistently exhibit near-isotropy, while larger scales inherit anisotropy from boundary conditions. This characterization highlights turbulence as a natural test bed for developing methods that address multiscale symmetries.

While we focus here on super-resolution, implicit augmentation from isotropy is a general property of turbulent statistics. Symmetry-aware methods have already been applied to other fluids tasks, such as wall-shear estimation and turbulence closures, where they yield gains in physical consistency under frame changes \citep{SukEtAl_2024,PoffEtAl_2025}. These successes suggest that the benefits of symmetry extend well beyond reconstruction tasks. Future work will investigate how the local inductive bias and translation equivariance of CNNs influence their ability to capture multiscale isotropy in turbulent flows. More broadly, our study underscores the importance of reasoning about distributional symmetries in the data itself, alongside architectural design, as a key ingredient for effective and physically consistent learning.

\begin{ack}
We acknowledge the support of the National Science Foundation under Cooperative Agreement PHY-2019786 (The NSF AI Institute for Artificial Intelligence and Fundamental Interactions). Julia Balla was supported by the Department of Defense (DoD) through the National Defense Science \& Engineering Graduate (NDSEG) Fellowship Program. Jeremiah Bailey was supported by the MIT Summer Research Program (MSRP). Elyssa Hofgard was supported by the U.S. Department of
Energy, Office of Science, Office of Advanced Scientific
Computing Research, Department of Energy Computational
Science Graduate Fellowship under Award Number DESC0024386. Ryley McConkey was supported by the Natural Sciences and Engineering Research Council of Canada (NSERC), Thornton Family Fund.
\end{ack}

\bibliographystyle{unsrtnat}
\bibliography{references}


\appendix

\section{Statistical homogenity and isotropy in turbulence}\label{app:definitions}

In turbulent flows, the velocity field $\mathbf{U}(\mathbf{x}, t)$ is a time-dependent random vector field, and its symmetries are naturally expressed in terms of distributions. For a vector field $\mathbf{U}:\Omega\!\to\!\mathbb{R}^d$, a group $G$ acts on both coordinates and components as $(\rho(g)\mathbf{U})(\mathbf{x}) \;=\; g\,\mathbf{U}\!\left(g^{-1}\mathbf{x}\right)$. 
The field is said to be \textit{statistically homogeneous} if all statistics are invariant under shifts in the origin of the coordinate system. Formally, for any translation $\mathbf{r}\in\mathbb{R}^d$ and any $N$, the $N$-point joint distribution is invariant: 
\[
\left(\mathbf{U}(\mathbf{x}_1, t), \ldots, \mathbf{U}(\mathbf{x}_N), t\right) \stackrel{d}{=} \left(\mathbf{U}(\mathbf{x}_1 + \mathbf{r}, t), \ldots, \mathbf{U}(\mathbf{x}_N + \mathbf{r}, t)\right).
\]
If, in addition, the distribution is invariant under rotations and reflections, the field is \textit{statistically isotropic}, i.e., for any rotation $g \in O(3)$, 
\[(\mathbf{U}(\mathbf{x}_1, t), \ldots, \mathbf{U}(\mathbf{x}_N, t)) \stackrel{d}{=} (g\mathbf{U}(g^{-1}\mathbf{x}_1, t), \ldots, g\mathbf{U}(g^{-1}\mathbf{x}_N, t)).
\]
Even in anisotropic flows, Kolmogorov’s local isotropy hypothesis posits that small-scale motions recover statistical isotropy when the Reynolds number---the ratio of inertial to viscous forces---is sufficiently high.

\section{Training hyperparameters}\label{app:training_params}

We use a fixed CNN super-resolution architecture for all of our experiments. The network contains two successive upsampling layers, each enlarging the input by a factor of $2$, resulting in an overall scale factor of $s=4$. 3D Convolutions use kernels of size $3$ kernels with reflection padding of one pixel on each side, followed by ReLU activations. All hidden layers have 128 channels. Models are trained with the Adam optimizer with learning rate $3 \times 10^{-4}$ and batch size $16$.

\section{Continuous vs. discrete rotational symmetry in the Navier-Stokes equations}\label{app:rotation_symmetry}
All evaluations in this work are carried out with respect to the octahedral group $O$ (rotations without inversions), as discretization and downsampling on a Cartesian grid breaks continuous rotational symmetry. In a continuum, the Navier–Stokes equations are invariant under the full 3D rotation group $SO(3)$. However, discretization on a Cartesian grid breaks this invariance, leaving only the rotational symmetries of the cube. Filtering operations used to generate low-resolution inputs further introduce deviations, so the data contain small but unavoidable rotational artifacts. Although in principle it is possible to test equivariance with respect to the entire $SO(3)$ group using interpolation schemes, such approaches introduce additional complexity and ambiguity in the comparison. We leave both $SO(3)$ evaluations and the inclusion of mirror and inversion symmetries for future work.

\end{document}